\documentclass[10pt,twocolumn,twoside]{IEEEtran} % For two columns

\usepackage{amsfonts}
\usepackage{amssymb}
\usepackage{amsmath}
\usepackage{amsthm}
\usepackage{graphicx}
\usepackage{subfigure}
\usepackage[verbose,nospace,sort]{cite}
\usepackage{algpseudocode}
\usepackage{algorithm}
\begin{document}
\newcommand{\xB}{\mathbf{B}}
\newcommand{\xE}{\text{E}}
\newcommand{\xe}{\mathbf{e}}
\newcommand{\xGF}{\text{GF}}
\newcommand{\xtr}{\text{tr}}
\newcommand{\xd}{\mathbf{d}}
\newcommand{\xH}{\mathbf{H}}
\newcommand{\xI}{\mathbf{I}}
\newcommand{\xtI}{\text{I}}
\newcommand{\xtlog}{\text{log}}
\newcommand{\xQ}{\mathbf{Q}}
\newcommand{\xP}{\mathbf{P}}
\newcommand{\xp}{\mathbf{p}}
\newcommand{\xR}{\mathbf{R}}
\newcommand{\xSF}{\text{SF}}
\newcommand{\xu}{\mathbf{u}}
\newcommand{\xU}{\mathbf{U}}
\newcommand{\xv}{\mathbf{v}}
\newcommand{\xV}{\mathbf{V}}
\newcommand{\xy}{\mathbf{y}}
\newcommand{\xX}{\mathbf{X}}
\newcommand{\xx}{\mathbf{x}}
\newcommand{\xz}{\mathbf{z}}
%\newcommand{\xdeg}{\textrm{deg}}

%----------------------------------------------------------------------
% Title Information, Abstract and Keywords
%----------------------------------------------------------------------
\title{Sub-Stream Fairness and Numerical Correctness in MIMO Interference Channels}

\author{Cenk M. Yetis$^{\dagger}$, Yong Zeng$^{\ddagger}$, Kushal Anand$^{\ddagger}$, Yong Liang Guan$^{\ddagger}$, and Erry Gunawan$^{\ddagger}$
\\ $^{\dagger}$ Electrical and Electronics Engineering Department, Mevlana University, Turkey
\\ $^{\ddagger}$ School of Electrical and Electronic Engineering, Nanyang Technological University, Singapore
\\ cenkmyetis@yahoo.com, $\{\text{ze0003ng,kush0005}\}$@e.ntu.edu.sg, $\{\text{eylguan,egunawan}\}$@ntu.edu.sg
\thanks{The work of C. M. Yetis in part and Y. Zeng, K. Anand, Y. L. Guan, and E. Gunawan in full was supported by the Department of the Navy Grant N62909-12-1-7015 issued by Office of Naval Research Global.}
}

\maketitle
\begin{abstract}
Signal-to-interference plus noise ratio (SINR) and rate fairness in a system are substantial quality-of-service
(QoS) metrics. The acclaimed SINR maximization (max-SINR) algorithm does not achieve fairness between user's
streams, i.e., sub-stream fairness is not achieved. To this end, we propose a distributed power control
algorithm to render sub-stream fairness in the system. Sub-stream fairness is a less restrictive design metric
than stream fairness (i.e., fairness between all streams) thus sum-rate degradation is milder. Algorithmic
parameters can significantly differentiate the results of numerical algorithms. A complete picture for
comparison of algorithms can only be depicted by varying these parameters. For example, a predetermined
iteration number or a negligible increment in the sum-rate can be the stopping criteria of an algorithm. While
the distributed interference alignment (DIA) can reasonably achieve sub-stream fairness for the later, the
imbalance between sub-streams increases as the preset iteration number decreases. Thus comparison of max-SINR
and DIA with a low preset iteration number can only depict a part of the picture. We analyze such important
parameters and their effects on SINR and rate metrics to exhibit numerical correctness in executing the
benchmarks. Finally, we propose group filtering schemes that jointly design the streams of a user in contrast to
max-SINR scheme that designs each stream of a user separately.
\end{abstract}

\section{Introduction}
The prominent paper \cite{85} proposed a new technique that was coined interference alignment (IA), and IA was shown to
achieve the upper bound of non-interfering signaling dimensions in an interference channel (IC). Later, the limits of IA in
multiple-input multiple-output (MIMO) ICs were mainly identified in \cite{188} and in consecutive papers \cite{166,167}. The
numerical results of IA first appeared in \cite{55}, where authors proposed a distributed IA (DIA) algorithm. In \cite{55}, IC
extension of the conventional signal-to-interference plus noise ratio maximization (max-SINR) for point-to-point channels was
also proposed. Max-SINR could achieve higher sum-rate than DIA in the low to mid signal-to-noise ratio (SNR) regime.
Subsequently many papers proposed other techniques that offered improved sum-rates than DIA and max-SINR could achieve. Please
refer to \cite{204} and the references therein. However, all these techniques overlooked an important quality-of-service (QoS)
metric, sub-stream fairness, i.e., fairness between streams of a user.

Although the acclaimed max-SINR algorithm lacks art in its design approach, its sum-rate results in MIMO ICs is surprisingly
satisfactory. It is the curiosity to design the streams of a user (sub-streams) jointly, thus sub-streams are not considered
as interference on one another. In fact, numerical results of an algorithm can significantly deviate by varying its
algorithmic parameters, and in addition, different algorithms have different responses to algorithmic parameters. Therefore
carefully scanning these parameters is important to benchmark entirely. For example, the sum-rate gap between max-SINR and DIA
in low to mid SNR regime can be emphasized or both can be asserted as not achieving sub-stream fairness when screening of
parameters shortfall.

In this paper, we propose a distributed power control algorithm (DPCA) in an ad-hoc manner to recognize
sub-stream fairness in the system. Particularly, we initially obtain the beamforming vectors via conventional
schemes including max-SINR and DIA with even power distributions, and then apply our DPCA to ensure sub-stream
fairness with a slightly increased algorithmic load. In the paper, we also propose two new algorithms that
jointly design sub-streams. Finally, we briefly discuss important algorithmic parameters and their influences on
benchmarks.

\section{System Model}
We consider a $K$-user interference channel, where there are $K$ transmitters and receivers with $M_k$ and $N_k$
antennas at node $k$, respectively. A transmitter has $d_k$ streams to be delivered to its corresponding
receiver. This system can be modeled as $\xy_k=\sum_{l=1}^K\xH_{kl}\xx_l+\xz_k, ~ \forall
k\in\mathcal{K}\triangleq\{1,2,...,K\},$ where $\xy_k \textrm{ and } \xz_k$ are the $N_k\times 1$ received
signal vector and the zero mean unit variance circularly symmetric additive white Gaussian noise vector (AWGN)
at the $k^{th}$ receiver, respectively. $\xx_l$ is the $M_l\times 1$ signal vector transmitted from the $l^{th}$
transmitter and $\xH_{kl}$ is the $N_k\times M_l$ matrix of channel coefficients between the $l^{th}$
transmitter and the $k^{th}$ receiver. $\xE[||\xx_l||^2]=p_l$ is the power of the $l^{th}$ transmitter. The
transmitted signal from the $l^{th}$ user is $\xx_l=\xU_l\xd_l$, where $\xU_l$ is the $M_l\times d_l$ precoding
(beamforming) filter and $\xd_l$ is $d_l\times 1$ vector denoting the $d_l$ independently encoded streams
transmitted from the $l^{th}$ user. The $N_k\times d_k$ receiver matrix is denoted by $\xV_k$.

\section{Overview of Conventional and New Algorithms} \label{sec:Overview}
This section begins with an overview of three conventional algorithms, DIA, max-SINR, and minimization of sum of
mean square errors (min-sum-MSE). Later in the section, we introduce two schemes that jointly design the streams
of a user as opposed to max-SINR scheme that considers streams of a user as interfering and designs the
beamforming vectors of a user independently. In other words, our proposed schemes are based on group filtering,
and allow collaboration between beamforming vectors of a user as opposed to max-SINR algorithm that undergoes
intra-user interference disadvantage. In the next section, we also show that reckoning the streams of a user as
whether interfering or not causes insignificant difference in sum-rate performance of max-SINR.

\subsection{DIA}
Interference alignment (IA) packs multi-user interference in a space separate from the desired signal space \cite{85}. This
scheme achieves optimal degrees of freedom (DoF) in MIMO ICs. IA is feasible provided the following conditions satisfied

\begin{subequations}
\begin{align}
\xV_k^\dagger\xH_{kj}\xU_j&=\mathbf{0}_{d_k}, ~ \forall k\neq j,\label{eqn:IAcondition1}\\
\text{rank}\left(\xV_k^\dagger\xH_{kk}\xU_k\right)&=d_k, ~ \forall k \in \mathcal{K}, \label{eqn:IAcondition2} \end{align}
\end{subequations}
where $\mathbf{0}_{d_k}$ indicates the $d_k\times d_k$ zero matrix. The second condition \eqref{eqn:IAcondition2} is
automatically satisfied for MIMO channels without specific structures. In \cite{55}, a DIA algorithm was proposed based on
network duality. Particularly, transmit and receive beamforming vectors are updated iteratively to minimize (align) the
interference. The interference leakage at receiver $k$ can be defined as ${\text{IL}_k=\xtr(\xV_k^\dagger\xQ_k\xV_k)}$, where
$\xtr(\mathbf{A})$ denotes the trace of matrix $\mathbf{A}$ and ${\xQ_k=\sum_{j=1,j\neq
k}^K\frac{p_j}{d_j}\xH_{kj}\xU_j\xU_j^\dagger\xH_{kj}^\dagger}$ is the interference covariance matrix of user $k$. To minimize
the interference, $d_k$ eigenvectors corresponding to the $d_k$ smallest eigenvalues of the interference covariance matrix are
assigned to IA receive beamforming vectors $\xv_{k,l}=\mathcal{V}_l(\xQ_k), ~ l=1,\cdots,d_k$, where $\xv_{k,l}$ is the
$l^{th}$ receive beamforming vector (i.e, $l^{th}$ column of the receive filter) of user $k$, and $\mathcal{V}_l(\mathbf{A})$
denotes the $l^{th}$ eigenvector ($\xe_l$) corresponding to the $l^{th}$ eigenvalue ($\it e_l$) of matrix $\mathbf{A}$. At
each iteration of the algorithm, assume eigenvectors are sorted in increasing order of corresponding eigenvalues, $\it
e_1\leq\it e_2\cdots\leq\it e_{d_k}$, and assigned to beamforming vectors in the same order, i.e., $\xv_{k,l}=\xe_l, ~
l=1,\cdots,d_k$. Before the algorithm converges, the interference seen by the $l^{th}$ beamforming vector can be lower than
the $(l+1)^{th}$ beamforming vector, whereupon SINRs and rates of succeeding sub-streams can be lower than the formers.

\subsection{Max-SINR}
As known IA is DoF optimal, in other words it achieves the maximum multiplexing gain at high SNR
\begin{equation} \label{eqn:ShannonUserRate}
R_k=\xtlog_2\left|\xI_{d_k}+\frac{\xV_k^\dagger\xR_k\xV_k}{\xV_k^\dagger\xB_k\xV_k}\right|
=d_k\xtlog_2(p_k)+o\left(\xtlog_2(p_k)\right),
\end{equation}
where $\xB_k=\xQ_k+\xI_{N_k}$ and $\xI_{x}$ are the interference plus noise and $x\times x$ identity matrices, respectively,
$\xR_k=\frac{p_k}{d_k}\xH_{kk}\xU_k\xU_k^\dagger\xH_{kk}^\dagger$ and $R_k$ are the covariance matrix and rate of user $k$,
respectively, finally $d_k$ is the DoF (multiplexing gain) achieved by user $k$, hence DoF is an approximation to rate. In
fact sum-rate, $R_\text{sum}=\sum_{k=1}^KR_k$, is more valuable than the DoF metric in real life. IA aims to minimize the
interference, in other words the desired signal power is not considered. Ergo these reasons, foci of IA are minimizing the
interference and achieving the optimal DoF, its sum-rate can be improved vastly in the low to mid SNR regime. As yet the
optimal max-SINR scheme is not known for ICs, IC extension of the optimal max-SINR filter for a single stream and
point-to-point system shows compelling sum-rate performances in ICs. The results are surprising since max-SINR can still
perform higher than DIA for high number of users and streams per user \cite{195} although it has the intra-user interference
disadvantage. Basically, max-SINR maximizes the SINR of each stream separately
\begin{equation} \label{eqn:maxsinrfilter}
\xv_{k,l}=\frac{\xB^{-1}_{k,l}\xH_{kk}\xu_{k,l}}{||\xB^{-1}_{k,l}\xH_{kk}\xu_{k,l}||},
\end{equation}
where $\xB_{k,l}=\xQ_{k,l}+\xI_{N_k},
~\xQ_{k,l}=\sum_{j=1}^K\frac{p_j}{d_j}\xH_{kj}\xU_j\xU_j^\dagger\xH_{kj}^\dagger-\xR_{k,l}$, and
$\xR_{k,l}=\frac{p_k}{d_k}\xH_{kk}\xu_{k,l}\xu_{k,l}^\dagger\xH_{kk}^\dagger$ is the covariance matrix of $l^{th}$ stream of
the $k^{th}$ user. Note that $\xB_{k,l}$ contains the intra-user interference. The filter \eqref{eqn:maxsinrfilter} is optimum
if \textit{only} it were to maximize the SINR of the $l^{th}$ stream of the $k^{th}$ user
\begin{equation}\label{eqn:SINRSF}
\text{SINR}^\xSF_{k,l}=\frac{\xv_{k,l}^\dagger\xR_{k,l}\xv_{k,l}}{\xv_{k,l}^\dagger\xB_{k,l}\xv_{k,l}}.
\end{equation}
Here SF stands for separate filtering since max-SINR designs each stream of a user independently from one
another. Max-SINR is apparently sub-optimal even for a point-to-point system with multiple streams since the
streams are competing with each other. A better approach is to design beamforming vectors by allowing
cooperation between them \cite{176}. As observed in DIA, stream SINRs and rates can be unbalanced due to the
inherent competition between the streams.

\subsection{Min-Sum-MSE}
In \cite{204}, an algorithm that minimized the sum of mean-square errors (min-sum-MSE) of all users was proposed. Although the
objective function of min-sum-MSE does not aim SINR and rate fairness between streams, the simulation results show a rather
well achieved fairness in terms of these QoS metrics.

\subsection{GEVD}
As the name indicates, generalized eigenvalue decomposition (GEVD)
is the generalized version of eigenvalue decomposition
\begin{equation}\label{eqn:GEVDSingleStream}
\xR_{k,l}\xv_{k,l}=\lambda_{k,l}\xB_{k,l}\xv_{k,l}.
\end{equation}
Since the maximum value of \eqref{eqn:SINRSF} is equal to the largest generalized eigenvalue ($\lambda_{\text{max}}$) of
\eqref{eqn:GEVDSingleStream}, generalized eigenvector corresponding to $\lambda_{\text{max}}$ is assigned to beamforming
vector $\xv_{k,l}$. GEVD solution \eqref{eqn:GEVDSingleStream} and max-SINR filter \eqref{eqn:maxsinrfilter} are equivalent
for single stream per user systems. For multi-stream per user systems, GEVD solution \eqref{eqn:GEVDSingleStream} can be
extended straightforwardly
\begin{equation} \label{eqn:GEVDMultipleStreams}
\xR_{k}\xV_{k}=\boldsymbol{\lambda}_k\xB_k\xV_{k},
\end{equation}
where ${\boldsymbol{\lambda}_k=\text{diag}(\lambda_{k,1},\cdots,\lambda_{k,d_k},\cdots,\lambda_{k,N_k})}$ are the generalized
eigenvalues. Here note that $\xB_k$ does not contain intra-user interference. GEVD in matrix form
\eqref{eqn:GEVDMultipleStreams} is solved by assigning $d_k$ generalized eigenvectors corresponding to $d_k$ largest
generalized eigenvalues to beamforming matrix $\xV_k$. Comparing \eqref{eqn:GEVDSingleStream} and
\eqref{eqn:GEVDMultipleStreams}, we see that streams of a user are competing in max-SINR solution whereas they are
collaborating in GEVD solution. Hence GEVD falls into group filtering schemes whereas max-SINR fits under separate filtering.
GEVD is another bandwidth inefficient scheme since it orthogonalizes the desired ($\xR_k$) and interference plus noise
($\xB_k$) signal spaces. We remind that the sum-rate of max-SINR similarly diminishes as SNR or the number of streams per user
increases since it acknowledges intra-user interference.

$\xV_k$ obtained from the GEVD solution \eqref{eqn:GEVDMultipleStreams} is the optimal filter for the problem
$\underset{\xV_k}{\text{max}}~\xtr\bigg(\big(\xV_{k}^\dagger\xB_k\xV_{k}\big)^{-1}\big(\xV_{k}^\dagger\xR_{k}\xV_{k}\big)\bigg),$
which is an approximation to the trace quotient maximization problem
\begin{equation}\label{eqn:TQM}
\underset{\xV_k}{\text{max}}~\frac{\xtr(\xV_{k}^\dagger\xR_{k}\xV_{k})}{\xtr(\xV_{k}^\dagger\xB_k\xV_{k})}.
\end{equation}

The stream SINR of group filtering (GF) schemes can be defined as
$\text{SINR}^{\xGF}_{k,l}=\frac{\xv_{k,l}^\dagger\xR_k\xv_{k,l}}{\xv_{k,l}^\dagger\xB_k\xv_{k,l}}.$ Hence
average SINR per user is given as a trace quotient
\begin{subequations}
\begin{align}
\overline{\text{SINR}}^\xGF_k&=\frac{1}{d_k}\sum_{l=1}^{d_k}\text{SINR}^{\xGF}_{k,l}=\frac{\xtr(\xV_{k}^\dagger\xR_{k}\xV_{k})}{\xtr(\xV_{k}^\dagger\xB_k\xV_{k})}&&
\label{eqn:AvSINR}\\
\intertext{provided beamforming vectors are adjusted to satisfy the condition}
&\hspace{1cm}\xV_{k}^\dagger\xB_k\xV_{k}=c\xI_{d_k},\label{eqn:GEVDcondition}
\end{align}
\end{subequations}
where $c$ is a constant. The motivation for maximizing average SINR, problem \eqref{eqn:TQM} given the condition
\eqref{eqn:GEVDcondition} holds, is given later in this section. Next we present a semidefinite programming
(SDP) scheme that optimally solves the average SINR maximization problem.

\subsection{SDP}
Trace quotient maximization arises in various fields of engineering, however we find only the SDP approach in \cite{197}
appropriate to use in the field of communications engineering. Next some comparisons are given. First, SDP, GEVD and max-SINR
come in order from the highest to the lowest average SINR achieving schemes. Second, while they all achieve proximate SINR per
user results, for single stream per user systems, they achieve the same results. Finally, it is worth mentioning the
convergence rate of SDP is much slower than min-sum-MSE, and the convergence rate of min-sum-MSE is much slower than the other
algorithms in the paper (DIA, max-SINR and GEVD). Further details and numerical results of SDP will be contained in journal
version.

\subsection{On Group Filtering}
As aforementioned, conventional max-SINR algorithm is sub-optimal since the SINR of each stream is maximized, thus the streams
of a user are considered as interference on one another. As highlighted, GEVD and SDP jointly optimize the transmit
beamforming vectors, thus the streams of a user are collaborating as opposed to the max-SINR algorithm. These two group
filtering based schemes can be used for maximization of average SINR, a useful QoS metric in communications. This metric
approximates the achievable rate at low SNR, and average SINR maximization simplifies the power allocation problem \cite{176},
where a centralized power allocation algorithm was proposed for MIMO downlink channels. In the next section, we also introduce
a DPCA based on average SINR maximization to achieve user fairness in MIMO ICs.

Shannon's formulation \eqref{eqn:ShannonUserRate} assumes a nonlinear decoder, in our context e.g., maximum
likelihood (ML) decoder \cite{177}. As well known, a nonlinear decoder jointly decodes the streams of a user,
thus in Shannon's formulation, the streams of a user are considered as non-interfering. Therefore, the
motivation of max-SINR to consider intra-user interference is not clear as shown next. Consider a modified
max-SINR filter that does not reckon intra-user interference, i.e., the interference covariance matrix $\xB_k$
is used instead of $\xB_{k,l}$ in \eqref{eqn:maxsinrfilter}. Our numerical results show that by using Shannon's
formula \eqref{eqn:ShannonUserRate}, there is nearly no difference between the modified and conventional
max-SINR algorithms. In other words, Shannon's formula absorbs the disparity between reckoning and not reckoning
the intra-user interference. Due to the same reason, separate and group filtering based schemes have similar
rate results when Shannon's formula is used.

\section{Sub-Stream Fairness}
In this section, we initially show the effects of stopping criteria on sub-stream fairness. Max-SINR algorithm dispenses
unequal stream SINRs and rates within each user, on the other hand DIA can provide sub-stream fairness depending on the
stopping criteria while min-sum-MSE is more robust to the stopping criteria. Before we propose a DPCA for balancing sub-stream
SINRs in MIMO ICs, we list some important algorithmic details next. As already stated, these details can significantly change
the perceptions on compared schemes. Further discussions on algorithmic parameters will be covered in the journal version.
\subsection{Algorithmic Parameters}
For all simulations in the paper, we fix the number of initializations of random transmit beamforming vectors to one. The
numerical results show that if more initializations are allowed, the sum-rate gap in the low to mid SNR regime between DIA and
max-SINR is reduced. The abbreviation \textit{Iter} in the plots stands for the number of iterations between uplink and
downlink, basically a downlink and then an uplink iteration count for two. Iter=$\emptyset$ indicates the number of iterations
is not predetermined, and the algorithm stops when the increment in the sum-rate is negligible
$|R_{\text{sum}}(n+2)-R_{\text{sum}}(n)|\leq \epsilon,$ where $n$ stands for the iteration number, and $\epsilon$ is $10^{-6}$
in our simulations. MC denotes the number of Monte Carlo simulations, and in the paper we present several results for the
system $K=3, M_k=M=N_k=N=4, d_k=d=2$. Finally, it is empirically observed for max-SINR that sum of stream rates
\begin{equation}\label{eqn:sumstream}
R_\text{sum-stream}=\sum_{k=1}^K\sum_{l=1}^{d_k}\xtlog_2(1+\text{SINR}_{k,l})
\end{equation}
approaches to Shannon's sum-rate $R_\text{sum}\approx R_\text{sum-stream}$ as the algorithm iterates \cite{208}.
In \eqref{eqn:sumstream} we do not indicate SINRs as SF or GF since the equations are valid for both. While for
GEVD the same approximation holds, for DIA and min-sum-MSE this approximation is looser as observed from
numerical results. Nevertheless, we evaluate the stream rates ($R_{k,l}$) as in \eqref{eqn:sumstream}, i.e.,
${R_{k,l}=\xtlog_2(1+\text{SINR}_{k,l})}$. Further discussion on this approximation will be included in the
journal version.

\begin{figure}[htb] \label{Fig:Fig1}
\centering \subfigure[Max-SINR, Iter=$\emptyset$] {
\includegraphics[height=5.25cm, width=9cm] {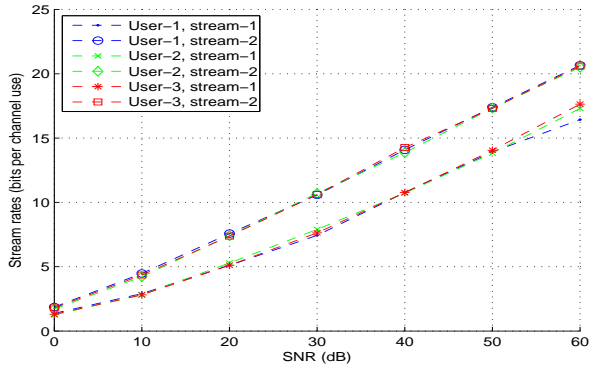} \label{Fig:Fig1a}
}
    \subfigure[Min-Sum-MSE, Iter=50] {
\includegraphics[height=5.25cm, width=9cm] {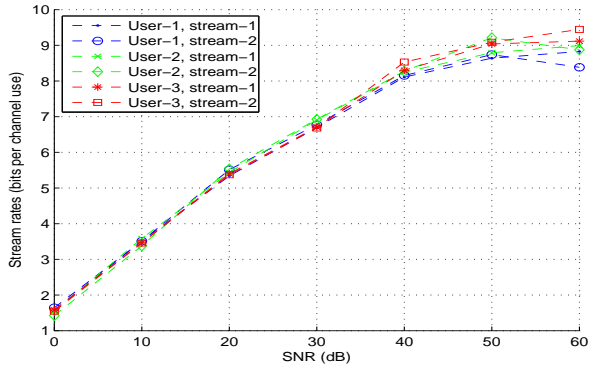} \label{Fig:Fig1b}
} \caption{Stream rates, MC=40.} \label{Fig:Fig1}
\end{figure}

In Fig. \ref{Fig:Fig1}, stream rates of max-SINR and min-sum-MSE are presented. While max-SINR appoints unequal rates to
streams of a user, min-sum-MSE allocates quite fair rates between all streams in the system. A similar observation can be
revealed for SINR results. From 0 to 60 dB, the sum-SINR ratios of the 2nd to the 1st streams,
$\sum_{k=1}^{K}(\text{SINR}_{k,2}/\text{SINR}_{k,1})$ are 5 and 1 in average for max-SINR and min-sum-MSE, respectively. As
the SNR increases to 60 dB, the imbalance of max-SINR reaches to 7 whereas it stays around 1 for min-sum-MSE. For all
mentioned algorithms, the balance improves or worsens in direct proportion to the iteration number. As seen in Fig.
\ref{Fig:Fig1a}, max-SINR cannot achieve fairness even for Iter=$\emptyset$, whereas min-sum-MSE still preserves fairness at a
much better level even for a low number of iteration as seen in Fig. \ref{Fig:Fig1b}. The simulation results of DIA show that
it achieves reasonable sub-stream fairness for Iter=$\emptyset$ while for fixed number of iterations, streams of a user are
imbalanced similar to max-SINR. The results of DIA are not plotted due to space limitations. As known, bit error rate (BER) is
influenced by the worst stream SINRs in the system. Thus, in general, min-sum-MSE can provide lower BER than max-SINR due to
its stream fairness, and max-SINR can achieve lower BER than DIA since it additionally aims to maximize desired signal power.
In \cite{204}, BER performances are compared for a small iteration number, Iter=16. Since the imbalance between streams in
max-SINR and DIA soar marginally more than min-sum-MSE as the iteration number decreases, BER gaps between these schemes are
significant at Iter=16. These findings again indicate the influence of iteration number, as other algorithmic details do, on
perceiving the complete picture.

\subsection{Power Control for Fairness}
SINR and rate fairness between users and streams can be achieved via joint power control and beamforming design or only via
power control design. In this section, we initially propose a DPCA based on average SINR metric, i.e., trace quotient
maximization with the constraint \eqref{eqn:GEVDcondition}. As alleged before, this is a handy metric that facilitates power
control problems. Pioneering works in power control mainly fall into two groups, centralized \cite{206} and decentralized
\cite{207} algorithms. A simple DPCA proposed in \cite{216} was extended to a more general framework in \cite{207}. In
\cite{207}, the author defined interference function as standard if it satisfied monotonicity and scalability properties, and
thereon constituted a standard power control algorithm (SPCA). Thanks to average SINR metric, SPCA becomes conveniently
available. Later in this section, we present another DPCA to achieve SINR fairness between sub-streams.

\subsubsection{User Fairness}
Recently, a SPCA for DIA was proposed in \cite{212}. The author erroneously defined SINR of a user as trace quotient
formulation \eqref{eqn:AvSINR}. Except this flaw in \cite{212}, the author successfully showed that the interference function
satisfied monotonicity and scalability properties. Here we only re-state the power constraint per user in its correct form
since the rest of the proof is similar to \cite{212}. User $k$ achieves Shannon's information rate under the condition
\begin{equation} \label{eqn:Shannoncondition}
\xtlog_2(1+d_k\overline{\text{SINR}}^\xGF_{k})\geq R_k \Rightarrow \overline{\text{SINR}}^\xGF_{k}\geq(2^{R _k}-1)/d_k,
\end{equation}
given \eqref{eqn:GEVDcondition} holds. By using \eqref{eqn:AvSINR} and \eqref{eqn:Shannoncondition}, the power constraint is
given as
$p_k\geq\xtI_k(\xp)=\frac{\xtr(\xV_{k}^\dagger\xB_k\xV_{k})}{\xtr(\xV_{k}^\dagger\xH_{kk}\xU_{k}\xU_{k}^\dagger\xH_{kk}^\dagger\xV_{k})},$
where $\xtI_k(\xp)$ is the interference function of user $k$ and $\xp=[p_1,\ldots,p_K]$ is the power vector of the system. We
note that the same author proposed a corrected version of the algorithm by introducing power control per stream in \cite{218}
that satisfied preset rate targets per user.

\subsubsection{Sub-Stream Fairness}
Fairness in the system can be achieved by two complementary approaches, maximization of minimum SINR subject to power
constraint or minimization of power subject to SINR constraint, and at three different levels, fairness between streams,
users, or sub-streams. Both problems achieve optimal solutions when, depending on the intended level, streams', users', or
sub-streams' SINRs are attained with equality \cite{217}. From more to less restrictive, stream, user and sub-stream fairness
come in order. Consequently, sub-stream fairness causes the least degradation in sum-rate, followed by user and stream
fairness. To this end, we propose an ad-hoc DPCA to retain fairness at the sub-stream level by using the later approach,
minimization of power subject to SINR constraint. Basically, transmit and receive beamforming vectors are initially obtained
via a scheme presented, but not limited to, in Section \ref{sec:Overview}. Then, in an ad-hoc manner, we apply the power
control presented in Algorithm \ref{alg:AdHocAlg}. The outer while loop searches for a feasible SINR target for each user.
Since there is a maximum power constraint, the optimal power values may not be feasible if SINRs are not well balanced before
power control applied. The superscripts denote iteration index in Algorithm \ref{alg:AdHocAlg},
$\xp_k^n=[p_{k,1}^n,\ldots,p_{k,d_k}^n]$ is the power vector of user $k$, and $p_{k,l}^n$ is the power for the $l^{th}$ stream
of the $k^{th}$ user at iteration $n$, $p_k^n=\sum_{l=1}^{d_k}p_{k,l}^n$. $\xB_k^{n-1}=\xQ_k^{n-1}+\xI_{N_k}$ and
$\xQ_k^{n-1}=\sum_{j=1,j\neq k}^K\frac{1}{d_j}\xH_{kj}\xU_j\xP_j^{n-1}\xU_j^\dagger\xH_{kj}^\dagger$ are interference plus
noise and interference covariance matrices, respectively,
$\xR_{k,l}^\prime=\xH_{kk}\xu_{k,l}\xu_{k,l}^\dagger\xH_{kk}^\dagger$ is akin to a covariance matrix,
$\xP_j^{n-1}=\text{diag}[p_{j,1}^{n-1},\ldots,p_{j,d_j}^{n-1}]$ is a diagonal matrix of sub-stream powers,
$\mathbf{1}=[1,\ldots,1]$ is all ones vector, $\boldsymbol{\delta}_k=[\delta_{k,1},\ldots,\delta_{k,d_k}]$ and
$\textbf{SINR}_k=\left[\text{SINR}_{k,1},\ldots,\text{SINR}_{k,d_k}\right]$ are the vectors of interference functions and
sub-stream SINRs, respectively.
\begin{algorithm}[htb!]
\footnotesize{\caption{Ad-Hoc DPCA} \label{alg:AdHocAlg}
\begin{algorithmic}[1]
\State Evaluate $\text{SINR}_{k,l}$ obtained from schemes presented in Section \ref{sec:Overview} \State
initialize $\text{SINR}_{k,l}^\prime=\text{SINR}_{k,l}$, $\forall k\in \mathcal{K}, \forall l\in\{1,\ldots,d_k\}
$\label{algsteps:InitSINRs} \State check=0 \While {check=$\sim$1} \State $\xp_k^0=\frac{p_k}{d_k}\mathbf{1}$,
$\xp_k^1=2\xp_k^0$, $\Gamma_k=\overline{\text{SINR}}_k^\prime$, $\forall k\in \mathcal{K}$
\label{algsteps:Initp} \State $n=1$ \While{$\sum_{k=1}^K||\xp_k^n-\xp_k^{n-1}||_1>\epsilon$} \State
$\delta_{k,l}=\frac{\xv_{k,l}^\dagger\xB_k^{n-1}\xv_{k,l}}{\xv_{k,l}^\dagger\xR_{k,l}^\prime\xv_{k,l}}$,
$\forall k\in \mathcal{K}, \forall l\in\{1,\ldots,d_k\} $ \State $x=2\max(\boldsymbol{\delta}_k)$, $p_k^T=0$,
$\forall k\in \mathcal{K}$ \For {counter=1:$d_k$}, $\forall k\in \mathcal{K}$ \State
$[\sim,y]=\min(\boldsymbol{\delta}_k)$ \State $p_{k,y}^n=\min(\Gamma_k\delta_{k,y},p_k-p_k^T)$ \State
$p_k^T=p_k^T+p_{k,y}^n$, $\delta_{k,y}=x$ \EndFor \State $n=n+1$\EndWhile \State Evaluate new SINRs
$\text{SINR}_{k,l}^\prime$ by using new power values $\xp_k^n$, $\forall k\in \mathcal{K}$ \If
{$\sum_{k=1}^K\sum_{\substack{m,n=1\\
m\neq n}}^{d_k}|\text{SINR}_{k,m}^\prime-\text{SINR}_{k,n}^\prime|\leq\epsilon$} \State check=1 \EndIf \EndWhile
\end{algorithmic}}
\end{algorithm}
\section{Numerical Results}
\begin{figure}[h!]
\centering
\includegraphics[height=5.25cm, width=9cm] {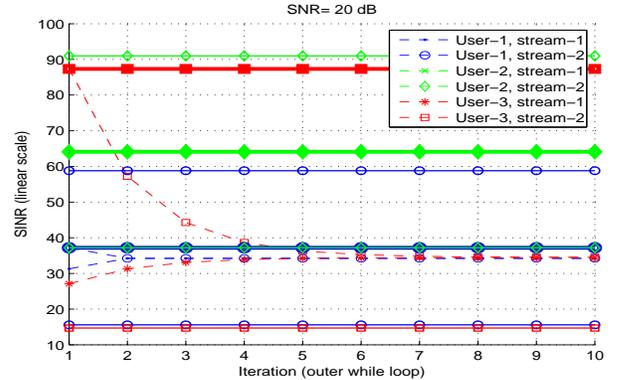}
\caption{Convergence behavior of the proposed power control in Algorithm \ref{alg:AdHocAlg}.} \label{Fig:Fig2}
\end{figure}

The proposed ad-hoc DPCA typically converges in a few outer while loop iterations, one such example for a channel realization
at 20 dB is given in Fig. \ref{Fig:Fig2}. Bold line in Fig. \ref{Fig:Fig2} indicates the average SINR and the other two
horizontal lines are the maximum and minimum SINR values of the user's streams before the power control algorithm is applied.
In Fig. \ref{Fig:Fig2}, the marks for horizontal lines are chosen the same with each corresponding user's mark of the second
stream. Note that the minimum SINR value of user 2 is close to the average SINR value of user 1, and the maximum SINR value of
user 3 that is around 160 is not plotted. The stream SINRs of user 2 achieve close to the average SINR value of user 2 from
the first iteration, thus the plots of these stream SINRs cannot be distinguished in the figure. As expected, the algorithm
maximizes the worst SINR of the sub-streams as seen in Fig. \ref{Fig:Fig2}, thus sub-stream fairness is achieved. Convergence
of the algorithm is guaranteed since a feasible point is searched in the outer while loop of the algorithm. While for user 1
and 2, it takes only 2 iterations and 1 iteration respectively, for user 3, it takes 23 iterations for sub-stream SINRs to
converge in fidelity of $\epsilon$. In Fig. \ref{Fig:Fig2}, only the first 10 iterations are plotted due to space constraints.
Note that the algorithm ends when the total sum of differences between sub-stream SINRs is negligible. While this approach
simplifies coding structure of Algorithm \ref{alg:AdHocAlg}, it causes redundant iterations for user 1 and 2.

\begin{figure}[htbp!]
 \centering
   \subfigure[with power control] {
\includegraphics[height=5.25cm, width=9cm] {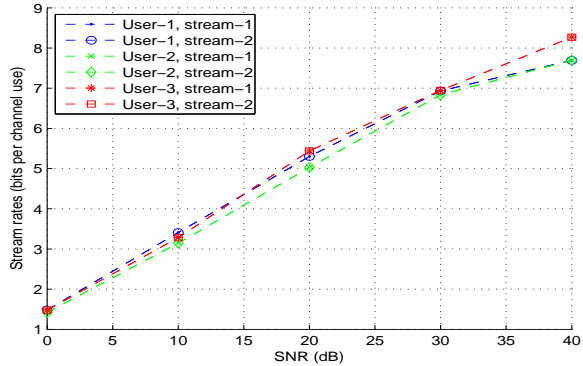} \label{Fig:Fig3a}
}
    \subfigure[without power control] {
\includegraphics[height=5.25cm, width=9cm] {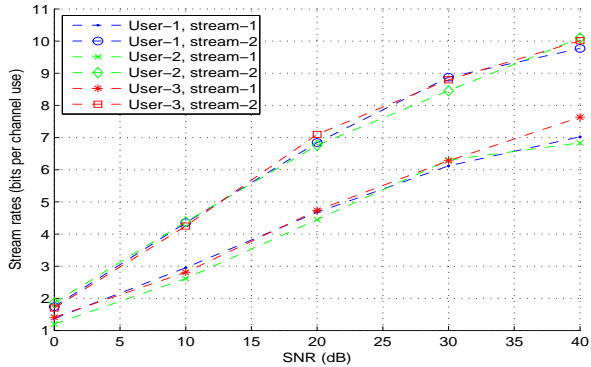} \label{Fig:Fig3b}
} \caption{Stream rates of max-SINR with and without the proposed power control in Algorithm \ref{alg:AdHocAlg}.}
\label{Fig:Fig4} \label{Fig:Fig3}
\end{figure}

In Fig. \ref{Fig:Fig3} and \ref{Fig:Fig4}, max-SINR without and with Algorithm \ref{alg:AdHocAlg} are compared for MC=20 and
Iter=50. User rates and sum-rates can be obtained from the stream rates, thus omitted. For the same system in Fig.
\ref{Fig:Fig3} and \ref{Fig:Fig4}, but with Iter=$\emptyset$, Algorithm \ref{alg:AdHocAlg} increased the simulation time by
only 20$\%$ (the run-time increased from 40 sec to 48 sec). Note that the SINR and rate results can be significantly improved
by designing beamforming and power vectors jointly.

\section{Conclusion}
Sub-stream fairness is an important QoS metric, and in addition it significantly influences the system's BER performance.
While conventional max-SINR scheme cannot achieve SINR fairness between sub-streams, DIA can achieve at a reasonable level
depending on the stopping criteria of the algorithm. To address this problem, we proposed an ad-hoc DPCA that retained
sub-stream fairness in the system with a slightly increased algorithmic load. We also proposed two new algorithms that
designed sub-streams jointly instead of independently as max-SINR did. Finally, we showed that numerical results and our
perceptions on benchmarks of schemes can be drastically shifted by varying algorithmic parameters. In addition to the future
research directions already pointed in the paper, BER analysis is another important research direction.

\begin{figure}[htbp!]
 \centering
   \subfigure[with power control] {
\includegraphics[height=5.25cm, width=9cm] {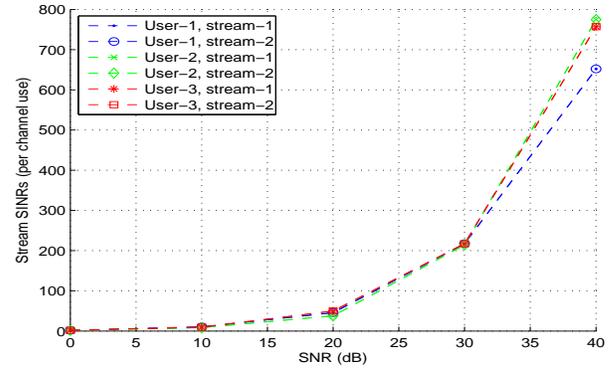} \label{Fig:Fig3c}
}
    \subfigure[without power control] {
\includegraphics[height=5.25cm, width=9cm] {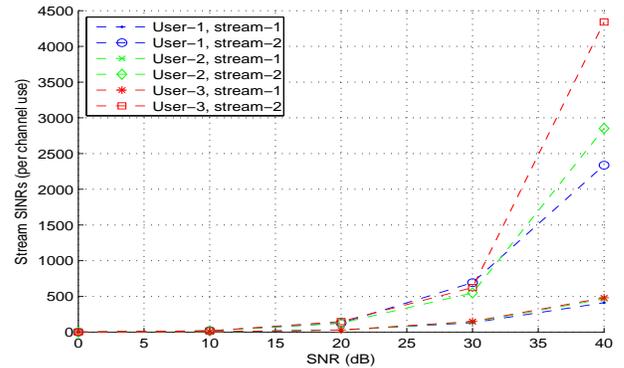} \label{Fig:Fig3d}
} \caption{Stream SINRs (linear scale) of max-SINR with and without the proposed power control in Algorithm
\ref{alg:AdHocAlg}.} \label{Fig:Fig4}
\end{figure}

\section{Acknowledgement}
We thank anonymous reviewers for helpful comments and thank the reviewer for bringing \cite{218} to our attention.

\bibliographystyle{IEEEtran}
\bibliography{IEEEabrv,IEEEfull}

\end{document}